\documentclass[aps,prl,twocolumn,showpacs]{revtex4}
\usepackage{graphicx}
\usepackage{dcolumn}

\begin{document}

\preprint{APS/123-QED}

\title{Evidence for Lattice Effects at the Charge-Ordering Transition in (TMTTF)$_2$X}

\author{M. de Souza$^{1}$}
\author{P. Foury-Leylekian$^{2}$}
\author{A. Moradpour$^{2}$}
\author{J.-P. Pouget$^{2}$}
\author{M. Lang$^{1}$}

\address{$^{1}$Physikalisches Institut, Goethe-Universit\"{a}t
Frankfurt, SFB/TRR 49, D-60438 Frankfurt(M), Germany}
\address{$^{2}$Laboratoire de Physique des Solides, Universit\'{e} Paris Sud, CNRS UMR 8502, Orsay, France }

\date{\today}

\begin{abstract}
High-resolution thermal expansion measurements have been performed
for exploring the mysterious "structureless transition" in
(TMTTF)$_{2}$X (X = PF$_{6}$ and AsF$_{6}$), where charge ordering
at $T_{CO}$ coincides with the onset of ferroelectric order.
Particularly distinct lattice effects are found at $T_{CO}$ in the
uniaxial expansivity along the interstack
$\textbf{\textit{c*}}$-direction. We propose a scheme involving a
charge modulation along the TMTTF stacks and its coupling to
displacements of the counteranions X$^{-}$. These anion shifts,
which lift the inversion symmetry enabling ferroelectric order to
develop, determine the 3D charge pattern without ambiguity.
Evidence is found for another anomaly for both materials at
$T_{int}$ $\simeq$ 0.6 $\cdot$ $T_{CO}$ indicative of a phase
transition related to the charge ordering.
\end{abstract}


\maketitle

Charge ordering (CO) has become a main issue for understanding
strongly correlated electron systems. Besides its diverse
manifestations in transition-metal oxides \cite{Imada 98}, it is
now recognized that the CO phenomena also play a key role for
$D_{2}A$ organic charge-transfer (CT) salts with a variety of
donor $D$ and acceptor $A$ molecules, such as (BEDT-TTF)$_{2}$X,
(TMTTF)$_{2}$X and (DI-DCNQI)$_{2}$Y (see \cite{Seo 06, Kakiuchi
07} and references therein). In these soft materials, the CO
induces an insulating state, which, by chemical substitution or
hydrostatic pressure, can be tuned towards a metallic and
low-temperature superconducting phase \cite{Jerome 04}. The CO
transition has been attributed to the importance of both on-site
$U$ and nearest-neighbor $V$ Coulomb interactions \cite{Seo 97}
with influence from electron-lattice coupling \cite{Riera 01, Clay
03, Brazovskii 03}. However, the definite role of the lattice for
charge localization has remained elusive.

Recently, an unexpected CO phase transition has been discovered
for the strongly one-dimensional (1D) (TMTTF)$_2$X salts,
preceding the transition into the tetramerized Spin-Peierls (SP)
ground state. At the transition temperature $T_{CO}$ $\simeq$
65\,K (X=PF$_6$) and 105\,K (AsF$_6$)  \cite{Chow 00} CO coincides
with the onset of ferroelectric order \cite{Nad 99,Monceau 01}.
This discovery, which shed new light on earlier observations on
related materials \cite{Laversanne 84}, together with the lack of
signatures in the magnetic properties, indicate that here one is
dealing with a surprising ferroelectric version of the
Mott-Hubbard state \cite{Monceau 01}. It has been argued that
effects of CO can be anticipated also in the more metallic
(TMTSF)$_{2}$X systems \cite{Monceau 01} and hence are relevant
over wide ranges in the unified phase diagram of (TM)$_2$X (TM =
TMTTF and TMTSF). The mystery surrounding the CO transition, now
known as "structureless transition", arose from the fact that up
until now no lattice effects at $T_{CO}$ have been observed
\cite{Pouget 96, Foury 02}. This is particularly puzzling as
atomic displacements, breaking the inversion symmetry, are
prerequisite for ferroelectricity to occur in these materials.

In this Letter we report for the first time lattice effects
accompanying the CO transition for (TMTTF)$_{2}$X by employing
thermal expansion measurements. The dilatometer used (built after
\cite{Pott 83}), has been particularly suitable for exploring
phase transitions in small and fragile crystals of organic CT
salts \cite{de Souza 07} due to its high resolution of $\Delta l/l
\sim 10^{-10}$, where $\textit{l}$ is the sample length.

Measurements have been performed on crystals of (TMTTF)$_2$X with
the centrosymmetric anions X = AsF$_6$ and PF$_6$. The crystals,
grown from THF using the standard procedure, are needle-shaped
with dimensions of about 10$\times$1$\times$0.5 mm$^{3}$ with the
needle axis parallel to the intrastack $\textbf{\textit{a}}$-axis.
Measurements were conducted along the $\textbf{\textit{a}}$-,
$\textbf{\textit{b'}}$- and $\textbf{\textit{c*}}$-axis
\cite{Note-AsF6}, where $\textbf{\textit{b'}}$ is perpendicular to
the $\textbf{\textit{a}}$-axis in the
($\textbf{\textit{a}}$,$\textbf{\textit{b}}$) plane and
$\textbf{\textit{c*}}$ is perpendicular to the
($\textbf{\textit{a}}$,$\textbf{\textit{b}}$) and
($\textbf{\textit{a}}$,$\textbf{\textit{b'}}$) planes. Care was
taken to keep the uniaxial stress, exerted by the dilatometer cell
on the crystal, below a maximum value of about 0.5 MPa.

\begin{figure}[floatfix]
\begin{center}
\vspace{-0.5cm}
  \includegraphics[width=0.8\columnwidth]{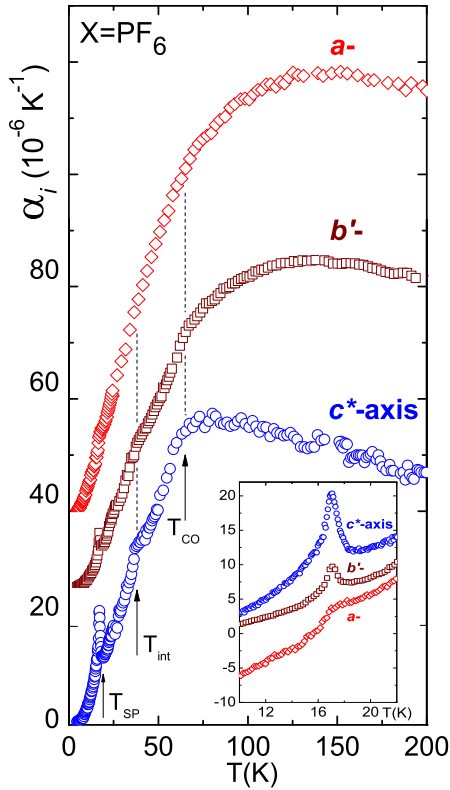}\\[-0.5cm]
   \caption{(Color online) Uniaxial expansivities along three orthogonal
   axes of (TMTTF)$_{2}$PF$_{6}$. The curves have been offset for clarity.
   Arrows mark the Spin-Peierls ($T_{SP}$) and charge-ordering ($T_{CO}$)
   transition temperatures reported in the literature and another transition ($T_{int}$) revealed here.
   Broken lines at $T_{CO}$ and $T_{int}$ are guides for the eyes.
   Inset shows details of the $\alpha_{i}$ anomalies at $T_{SP}$.}
 \label{fig:1}
\end{center}
\end{figure}

\begin{figure}[floatfix]
\begin{center}
\includegraphics[width=0.8\columnwidth]{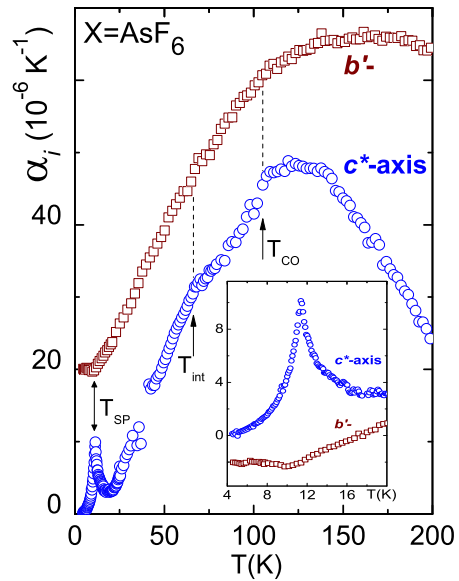}\\[-0.5cm]
\caption{(Color online) Uniaxial expansivities along the
$\textbf{\textit{b'}}$- and $\textbf{\textit{c*}}$-axis of
(TMTTF)$_{2}$AsF$_{6}$ with phase transition temperatures (arrows)
as defined in Fig.\,\ref{fig:1}. Gap in $\alpha_{c*}$ data between
37 and 41\,K correspond to range of large noise. Curves have been
offset for clarity. Broken lines are guides for the eyes. Inset
shows details of the anomalies at $T_{SP}$.} \label{fig:2}
\end{center}
\end{figure}

In Figs.\,\ref{fig:1} and \ref{fig:2} we show the results of the
uniaxial coefficient of thermal expansion $\alpha_{i}(T)=l_{i}
^{-1}dl_{i}/dT$ ($i$ = $\textit{a}$, $\textit{b'}$ or
$\textit{c*}$) for X = PF$_{6}$ and AsF$_{6}$ below 200 K
\cite{Note-PF6}. The data are dominated by large and anisotropic
lattice contributions. However, deviations from an ordinary
lattice expansion, characterized by an $\alpha(T)$ which
monotonously increases with $T$ (Debye-like), become evident at
higher temperatures. Here $\alpha_{i}$ decreases with increasing
temperature, indicating the action of a negative contribution
which grows with temperature. This effect is particularly strongly
pronounced in $\alpha_{c^{*}}$ for both salts and larger for the X
= AsF$_{6}$ when compared to the PF$_{6}$ salt.

Such a sizable negative contribution is unlikely to originate in
electronic degrees of freedom coupled to the lattice. Also a
magnetic contribution, $\alpha_{mag}$, due to 1D spin excitations
\cite{Dumm 00}, located at $k_{B}T^{\alpha} \simeq 0.48\cdot J$
\cite{Lang 04} ($J$ is the exchange coupling constant), is
unlikely as the sign of $\alpha_{mag}$ is given by the pressure
dependence of $J$ \cite{Bruehl 07}, which is positive here
\cite{note 3}. Rather, the negative contribution may indicate some
unusual lattice dynamics. As a possible mechanism, we propose that
rotational or translational modes of rigid PF$_{6}$ or AsF$_{6}$
units linked to the TMTTF molecules via $F$-$S$ halogen bonds
cause a negative contribution to $\alpha$. The size of this
contribution is expected to grow with the size of the anion.
Indeed, these octahedral anions, trapped in centrosymmetric
cavities delimited by the methyl groups \cite{Pouget 96}, are
known to be highly disordered at high temperatures and thought to
be rotating \cite{McBrierty 82}. As has been discussed in
connection with "negative thermal expansion" materials
\cite{negative alpha}, such "rigid-unit modes" may pull the entire
structure inwards giving rise to a lattice contraction on thermal
excitation.

In addition, the data in Figs.\,\ref{fig:1} and \ref{fig:2}
exhibit distinct, sharp features indicative of phase transitions.
Their positions coincide with the transition temperatures into the
CO and SP state (arrows labelled $T_{CO}$ and $T_{SP}$ in
Figs.\,\ref{fig:1} and \,\ref{fig:2}) reported in the literature,
e.g., Refs.\,\cite{Chow 00, Dumm 00}. Prominent effects show up at
$T_{SP}$ which are most strongly pronounced in $\alpha_{c^{*}}$
for both salts, cf.\,insets of Figs.\,\ref{fig:1} and
\,\ref{fig:2}. A detailed analysis of the phase transition
anomalies at $T_{SP}$ will be published elsewhere.

Clear signatures in $\alpha_{i}$ are also revealed at $T_{CO}$.
The lattice effects accompanying the CO transition constitute the
central results of this paper. The strongest effect is again
observed in $\alpha_{c^{*}}$, yielding a sharp kink at $T_{CO}$
followed by a rapid reduction immediately below the transition
temperature. The anomaly is similar for both salts,
cf.\,Figs.\,\ref{fig:1} and \ref{fig:2}, albeit somewhat more
distinct for the AsF$_{6}$ system. A corresponding feature, though
less strongly pronounced, can be seen also in $\alpha_{b'}$ for X
= PF$_{6}$, whereas it is less evident in $\alpha_{b'}$ for X =
AsF$_{6}$. A still smaller, if any, effect at $T_{CO}$ is found in
$\alpha_{a}$ for PF$_{6}$. We stress that measurements of
$\alpha_{c^{*}}$ on a second X = PF$_{6}$ crystal yielded
practically identical results to those shown in Fig.\,\ref{fig:1}.
The anomalous $T$-dependences of $\alpha_{c^{*}}$ in
Figs.\,\ref{fig:1} and \ref{fig:2} suggest a relation between
$T_{CO}$ and the negative thermal expansion contribution: upon
cooling through $T_{CO}$, this negative contribution vanishes
giving way to a positive slope $d\alpha_{c^{*}}/dT > 0$ for $T <
T_{CO}$.

The data in Figs.\,\ref{fig:1} and \,\ref{fig:2} disclose yet
another anomaly indicative of a phase transition which has been
overlooked so far: at an intermediate temperature $T_{int}$
$\simeq$ (39 $\pm$ 2)\,K (PF$_{6}$) and (65 $\pm$ 2)\,K
(AsF$_{6}$), the $\alpha_{i}$ data for both salts reveal a sharp
kink, which is most strongly pronounced in $\alpha_{c^{*}}$.

\begin{figure}
\includegraphics[width=\columnwidth]{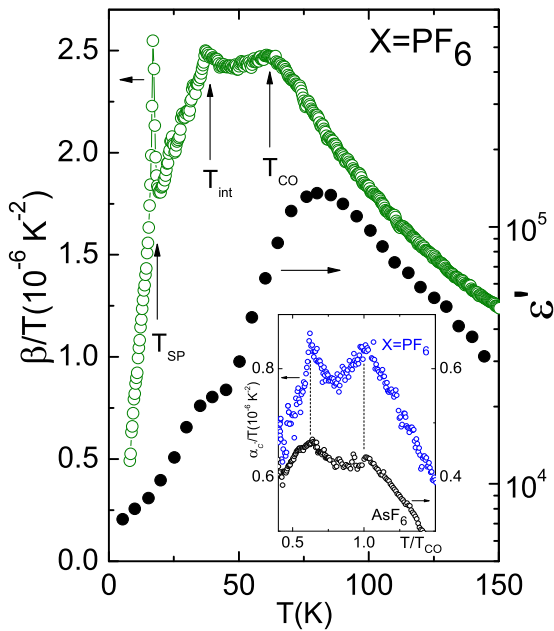}\\[-0.5cm]
\caption{(Color online) Left scale: Volume expansivity $\beta =
\alpha_{a} + \alpha_{b'} + \alpha_{c^{*}}$ (open circles) as
$\beta/T$ for (TMTTF)$_{2}$PF$_{6}$ determined from the data in
Fig.\,\ref{fig:1}. Arrows at $T_{SP}$, $T_{CO}$ and $T_{int}$ are
defined in Fig.\,\ref{fig:1}. Right scale: dielectric permittivity
$\epsilon'$ reproduced from Ref.\,\cite{Nad 06} plotted on the
same $T$ scale. Inset: $\alpha_{c^{*}}$ data for (TMTTF)$_{2}$X
with X = PF$_{6}$ (left scale) and X = AsF$_{6}$ (right scale) as
$\alpha_{c^{*}}/T$ vs.\,$T/T_{CO}$.} \label{fig:3}
\end{figure}

The various anomalies become particularly clear in the volume
expansion coefficient $\beta = \alpha_{a} + \alpha_{b'} +
\alpha_{c^{*}}$, shown in Fig.\,\ref{fig:3} for the X = PF$_{6}$
salt as $\beta/T$ vs.\,$\textit{T}$. The data unveil striking
similarities in the anomalies at $T_{CO}$ and $T_{int}$, i.e.\,,
sharp kinks accompanied by distinct changes in the slope,
indicative of a common nature of both transitions. In fact, an
origin of $T_{int}$ related to CO is corroborated by reexamining
results of the dielectric permittivity $\epsilon'$ for the same
salt \cite{Monceau 01} also shown in Fig.\,\ref{fig:3}. The sharp
kink in $\beta/T$ at $T_{int}$ coincides with a second peak in
$\epsilon'$ lying on the low-$T$ side of the main $\epsilon'$
maximum. We emphasize that the shift in the position of the latter
relative to the feature in $\beta$ is likely due to the frequency
dependence in $\epsilon'$ observed for this salt \cite{Nad 99}. In
the inset of Fig.\,\ref{fig:3}, we compare the expansivity results
for the X = PF$_{6}$ with those for the AsF$_{6}$ salt by plotting
both data sets on a reduced temperature scale $T/T_{CO}$. Due to
the lack of $\alpha_{a}$ data for the AsF$_{6}$ salt
\cite{Note-AsF6}, the comparison is made for $\alpha_{c^{*}}$
where the effects are most strongly pronounced. The coincidence in
the peak positions, implying $T_{int}$ to scale with $T_{CO}$ for
both compounds, suggests that these two features are closely
related to each other.

\begin{figure}
\includegraphics[width=0.7\columnwidth]{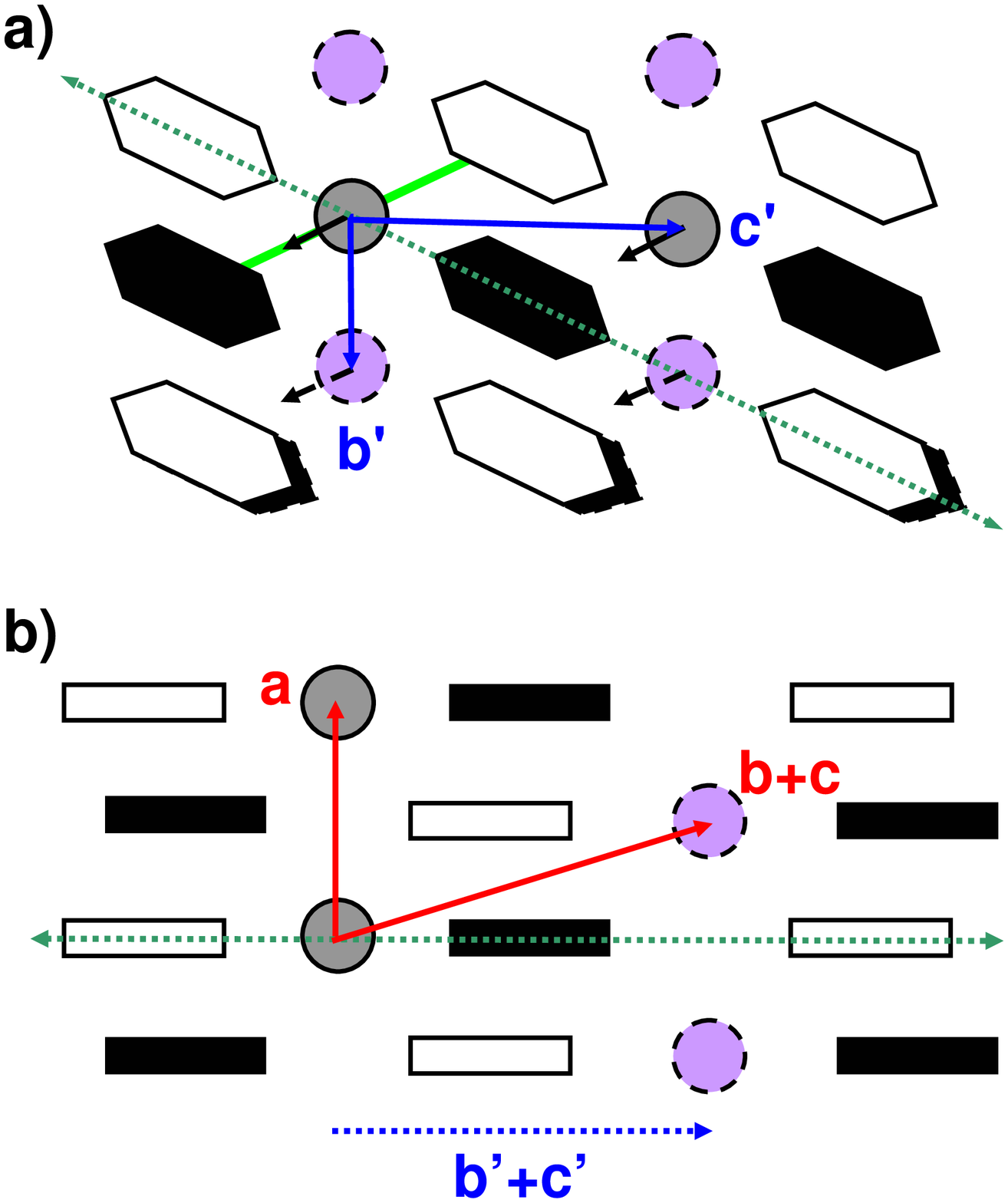}\\[-0.5cm]
\caption{(Color online) a) Schematic arrangement of TMTTF
molecules (hexagons/rectangles) and anions (circles) in the
($\textbf{\textit{b'}}$,$\textbf{\textit{c'}}$) plane, i.e.,
perpendicular to the intrastack $\textbf{\textit{a}}$-axis, and b)
the ($\textbf{\textit{a}}$,$\textbf{\textit{b'}} +
\textbf{\textit{c'}}$) plane. Green lines exemplarily show short
$S$-$F$ contacts. Black (white) hexagons/rectangles indicate
charge $\rho$ = 0.5 + $\delta$ (0.5 $-$ $\delta$) on the TMTTF
molecule in the CO state. Closed and dotted symbols refer to
positions within the
($\textbf{\textit{b'}}$,$\textbf{\textit{c'}}$) plane and shifted
by about $\textbf{\textit{a}}$/2 along the
$\textbf{\textit{a}}$-axis, respectively. Arrows indicate the
proposed shifts of the anions X$^{-}$ towards the (TMTTF)$^{\rho =
0.5 + \delta}$ molecules. } \label{fig:4}
\end{figure}

The uniaxial expansivity data in Figs.\,\ref{fig:1} and
\ref{fig:2} provide clear evidence that it is the interstack
$\textbf{\textit{c*}}$-axis which is most strongly involved in the
transition at $T_{CO}$. The $\textbf{\textit{c*}}$-direction is
distinct in that it incorporates the anions X: along
$\textbf{\textit{c*}}$,
($\textbf{\textit{a}}$,$\textbf{\textit{b}}$) planes of TMTTF
molecules alternate with planes of anions X,
cf.\,Fig.\,\ref{fig:4}. By contrast, the
$\textbf{\textit{a}}$-axis lattice parameter, which is determined
by intrastack interactions between adjacent TMTTF molecules,
remains practically unaffected by the transition at $T_{CO}$. It
therefore seems reasonable to include the anions and their
coupling to the TMTTF molecules in the discussion of the CO
process. In Fig.\,\ref{fig:4} we propose a simple scheme, which is
to involve the charge degrees of freedom on the TMTTF stacks and
their coupling to the anions. Upon cooling through $T_{CO}$, the
charge $\rho$ on the TMTTF molecule changes from a homogeneous
distribution with $\rho$ = 0.5 (in units of $e$) above $T_{CO}$ to
a modulated structure which alternates by $\pm \delta$ along the
TMTTF stacks below. For deriving the resulting 3D charge pattern,
we start by considering a stack of anions along the
$\textbf{\textit{a}}$-axis and the two nearest-neighbor stacks of
TMTTF molecules linked via short $S$-$F$ contacts (green lines in
Fig.\,\ref{fig:4}). For a fixed charge modulation on one of the
stacks, the electrostatic energy of the whole array can be reduced
if one of the anions' nearest-neighbor TMTTF molecules is charge
rich (TMTTF)$^{(\rho_{0} + \delta)}$ (black symbols), the other
one charge poor (TMTTF)$^{(\rho_{0} - \delta)}$ (white symbols),
while the anions perform slight shifts towards the charge rich
ones. The resulting anion displacements (indicated by the arrows
in Fig.\,\ref{fig:4}), which are uniform for all anions and lift
the inversion symmetry, together with the minimization of Coulomb
energies of adjacent stacks along the $\textbf{\textit{b}}$ axis
determine the 3D charge pattern unambiguously.

The importance of the anion potential for stabilizing the CO state
in the (TMTTF)$_{2}$X salts has been pointed out by several
authors \cite{Monceau 01, Riera 01, Brazovskii 03, Yu 04, Pouget
06}. In Ref.\,\cite{Riera 01} it was argued that for small Peierls
couplings, a sufficiently strong coupling to the anion
displacement field can generate a $4k_{F}$ CO state accompanied by
a uniform $\textbf{\textit{q}}$ = $(q_{\parallel}, q_{\perp})$ =
(0, 0) displacement of the anions with respect to their symmetric
positions. The latter state is consistent with the 3D displacement
pattern proposed here.

The anomalies in $\alpha_{i}$ disclosed at $T_{CO}$ are small and
lack any indications for discontinuous changes of the lattice
parameters, consistent with a second-order phase transition. A
detailed analysis of the phase transition anomaly is, however,
precluded by the anomalous background expansivity, although, the
rapid reduction of $\alpha_{c^{*}}$ immediately below $T_{CO}$
indicates a slightly broadened step-like anomaly, i.e., a
mean-field type transition. This would in fact be expected at the
CO transition as a result of long-range Coulomb forces, and
clearly revealed by the Curie-Weiss dependence of $\epsilon'$
\cite{Monceau 01}.

Yet, as pointed out above, the CO transition appears to
significantly affect the overall $\textbf{\textit{c*}}$-axis
expansivity, as $T_{CO}$ roughly coincides with the temperature
below which the negative contribution to $\alpha_{c^{*}}$ is no
longer active. The process involved may be appreciated within the
"rigid-unit mode" scenario proposed here to account for this
negative $\alpha$ contribution: above $T_{CO}$ CO fluctuations,
evident from the dielectric measurements \cite{Monceau 01} to
persist up to high temperatures, cause, via $S$-$F$ contacts,
positional fluctuations of the anions towards their new off-center
equilibrium positions. These positional fluctuations may provide
an effective damping of the anions' rigid-unit modes. Upon cooling
through $T_{CO}$, the CO becomes static, giving rise to a freezing
of these modes and, with it, a disappearance of the negative
contribution to $\alpha$.

Finally, we comment on the peak anomaly at $T_{int}$ in $\beta/T$
which is likely a phase transition related to the CO state.
Possible scenarios may include disorder-related (relaxor) effects
(either at $T_{CO}$, $T_{int}$ or both), or two consecutive
transitions associated with different CO patterns. Arguments in
favor of the latter possibility can be derived from infrared
spectra for both salts \cite{Dumm 05} suggesting qualitative
changes in the CO state below $T_{int}$, and the sharpness of the
anomalies in $\alpha_{c^{*}}$ observed here. On the other hand,
ferroelectricity arising from a neutral-ionic type of transition
akin to that discussed for TTF-chloranil \cite{Hubbard 81}, cannot
be ruled out. Here the transition occurs over a wide temperature
range with an onset expected to be a third-order transition
\cite{Hubbard 81}.

In conclusion, the mysterious CO transition coupled to
ferroelectricity in (TMTTF)$_{2}$PF$_{6}$ and
(TMTTF)$_{2}$AsF$_{6}$ has been explored by using high-resolution
thermal expansion measurements. The study reveals for the first
time evidence for lattice effects accompanying the transition,
thereby solving the long-standing puzzle surrounding this
hitherto-called \textit{structureless} transition. Based on the
directional dependence of the observed effects, we propose a
scheme involving the CO along the TMTTF stacks and its coupling to
anion X$^{-}$ displacements. The proposed uniform anion shifts,
locking the charge modulation of adjacent TMTTF stacks, determine
the 3D charge pattern unambiguously. The mechanism involved, being
similar to the scenario discussed for the Wigner crystallization
in (DI-DCNQI)$_{2}$Ag \cite{Kakiuchi 07}, highlights the intricate
role of the lattice degrees of freedom for stabilizing a CO ground
state in a wide class of materials.


\begin{thebibliography}{}
\bibitem{Note-AsF6} Measurements along the $\textbf{\textit{a}}$-axis for X = AsF$_6$ failed because the crystals
cleaved as soon as a small stress was exerted by the dilatometer -
an effect which is assigned to the highly 1D character of this salt.
\bibitem{Note-PF6} In \cite{Lang 04} results on an interstack expansion
coefficient on a different X = PF$_{6}$ crystal were given. The
present results on well-oriented crystals suggest the data in
Fig.\,1 of \cite{Lang 04} to be composed of contributions from
$\alpha_{b'}$ and $\alpha_{c^{*}}$, indicating some misalignment
of the crystal studied there.
\end{thebibliography}
\end{document}